\documentclass[journal]{IEEEtran}

\usepackage{cite}
\usepackage{amsmath}
\usepackage{algorithmic}
\usepackage{array}
\usepackage{xcolor}
\usepackage{url}
\usepackage{etoolbox} % for '\patchcmd' macro
\patchcmd{\subequations}{{1}}{{-1}}{}{}  % decrement the equation counter
\patchcmd{\subequations}{\alph}{.\arabic}{}{} % change display format of eq. counter
\usepackage{tabularx} %for automation of column widths so that the table spans the width defined by textwidth and linewidth

\ifCLASSINFOpdf
\usepackage[pdftex]{graphicx}
\else
 \usepackage[dvips]{graphicx}
\fi

\begin{document}

\title{  \textcolor{blue}{\underline{\Large{ Preprint of a manuscript  } } } \newline
	DEEP LEARNING BASED FORECASTING OF INDIAN SUMMER MONSOON RAINFALL}

\author{Bipin Kumar, Namit Abhishek,  Rajib Chattopadhyay,  Sandeep George, Bhupendra Bahadur Singh, Arya Samanta, B.S.V.  Patnaik, Sukhpal Singh Gill, Ravi S. Nanjundiah and Manmeet Singh
\thanks{Bipin Kumar, Rajib Chattopadhyay, Bhupendra Bahadur Singh, Ravi S. Nanjundiah and Manmeet Singh are with Indian Institute of Tropical Meteorology, Ministry of Earth Sciences, Dr. Homi Bhabha Road, Pune, 411008, India. }
\thanks{Namit Abhishek and Arya Samantha are with Indian Institute of Science Education and Research, Dr. Homi Bhabha Road, Pune, 411008, India}
\thanks{Sandeep George and B.S.V. Patnaik, are with Indian Institute of Technology, Madras, Chennai, 600036, India.}
\thanks{Sukhpal Singh Gill is with Queen Mary University of London
	Mile End Road, London E1 4NS, UK.}
\thanks{Manmeet Singh is also with IDP in Climate Studies, Indian Institute of Technology, Bombay, Powai, Mumbai, 400076, India}
\thanks{Rajib Chattopadhyay is also with CRS, India Meteorological Department, Pune}
\thanks{ Ravi S Nanjundiah is also with CAOS, Indian Institute of Science, Bangalure,560012, India.}
}

% The paper headers
%\markboth{Journal of \LaTeX\ Class Files,~Vol.~14, No.~8, August~2015}%
%{Shell \MakeLowercase{\textit{et al.}}: Bare Demo of IEEEtran.cls for IEEE Journals}

% make the title area
\maketitle

% As a general rule, do not put math, special symbols or citations
% in the abstract or keywords.
\begin{abstract}
Accurate short-range weather forecasting has significant implications for various sectors. Machine learning based approaches, e.g. deep learning, have gained popularity in this domain where the existing numerical weather prediction (NWP) models still have modest skill after a few days. Here we use a ConvLSTM network to develop a deep learning model for precipitation forecasting. The crux of the idea is to develop a forecasting model which involves convolution based feature selection and uses long term memory in the meteorological fields in conjunction with gradient based learning algorithm. Prior to using the input data, we explore various techniques to overcome dataset difficulties. We follow a strategic approach to deal with missing values and discuss the model’s fidelity to capture realistic precipitation. The model resolution used is (~25 km). A comparison between 5 years of predicted data and corresponding observational records for 2 days lead time forecast show correlation coefficients of 0.67 and 0.42 for lead day 1 and 2 respectively. The patterns indicate higher correlation over the Western Ghats and Monsoon trough region (~0.8 and ~0.6 for lead day 1 and 2 respectively). Further, the model performance is evaluated based on skill scores, Mean Square Error, correlation coefficient and ROC curves. This study demonstrates that the adopted deep learning approach based only on a single precipitation variable, has a reasonable skill in the short range. Incorporating multivariable based deep learning has the potential to match or even better the short range precipitation forecasts based on the state-of-the-art NWP models.
\end{abstract}

% Note that keywords are not normally used for peerreview papers.

\IEEEpeerreviewmaketitle

\section{Introduction}

\IEEEPARstart{F}{or} the two billion-plus residents of South Asia, monsoon precipitation is vital and directly impacts the economy and resources of the region. In particular, monsoon is considered the bread and butter of a large section of Indian society and is linked to the reversal of the wind patterns between winter and summer season. The period spanning June to September months, is considered the summer monsoon season for the sub-continent as most of the precipitation occurs during this interval. More robust classification of monsoonal regions is delineated based on the precipitation rate where the local summer-minus-winter precipitation rate exceeds 2.5 mm/day \cite{wang2011diagnostic}. The Indian subcontinent lies at the centre of the monsoonal region in South Asia surrounded by the Indian Ocean. With a standard deviation of sub 10\%, the interannual variability of the Indian Summer Monsoon (ISM) plays a major role in overall agricultural production \cite{rao2019monsoon, rajeevanindia,gadgil2003indian}.  

The studies on monsoon have been traditionally performed using numerical models of the weather and climate \cite{krishnan2020progress, singh2020fingerprint}, which solve partial differential equations of the atmosphere-ocean-land coupled systems. In the Indian context, the models focusing on different temporal scales of the Indian monsoon, viz., short-range to climate scale, are being used in research and operational mode to understand the monsoon better and disseminate the information to the stakeholders.  In recent times, the need for better forecasting has risen for several specific applications where the skills of dynamical models are still modest, owing to different global climate trends and climate change. In general, methods for predicting different meteorological variables use numerical weather prediction techniques by solving a set of higher-order non-linear differential equations. 

Short-range forecasting, i.e.,  1-3  days in advance, is very important, particularly,  in the context of the monsoon region as high-impact weather events are increasing with global warming \cite{goswami2006increasing}. An accurate assessment of the sub-district scale weather a few days ahead can arm the planners and administrators to take necessary measures in containing the potential damage. For example, usage of numerical weather prediction towards short-range precipitation forecasts can help in mitigating the impacts of cloud bursts, heavy-to-very-heavy extreme rainfall. Short-range prediction over India is being carried out by a suite of dynamical models; at present the highest spatial resolution of these models is $\approx$12.5km \cite{rajeevanindia, mukhopadhyay2019performance}. The improvements in short-range weather prediction have led to trickle-down benefits in various other sectors as well such as aviation, health, etc. Having shown tremendous progress in the last decade, such models sometimes fail to capture extreme rainfall events. For example, the National Centers for Environmental Prediction (NCEP) based Global Forecast System (GFS) T1534 ($\approx$12.5 km), has shown significant improvements in the short range operational forecasts over India \cite{mukhopadhyay2019performance}. However, even such  an advanced model underestimates the heavy to very heavy rainfall while the extremely heavy rainfall categories are only better at       the shorter lead times. There could be various reasons for these issues, particularly in India, such as the complex, non-linear and turbulent weather in the tropical regions and the usage of parameterization schemes generating precipitation in the model.  Other than numerical models, statistical and feature selection-based Artificial Neural Network (ANN) models have been used in the past to predict rainfall in different time scales with some success \cite{saha2016autoencoder, dasgupta2020exploring} . These models employ two popular concepts: feature selection and then prediction using statistical or simple machine learning algorithms \cite{saha2016autoencoder,kim1999eof, goswami2003potential, chattopadhyay2008objective}.

In the last decade, deep learning has emerged as a potential methodology to solve complex, non-linear problems by un-wrapping the nonlinearities in different layers of the deep neural network \cite{zeiler2014visualizing} . Such developments have occurred due to the availability of hardware capable of performing memory-intensive convolution operations that were not possible when neural convolution networks were first proposed \cite{lecun1998gradient} . Moreover, the development of software stacks such as open-source libraries (Tensor Flow, PyTorch, and others) has led to removing barriers to learning the field. The non-linear operators that have gained prominence in the Computer Vision community can be applied to weather and climate science problems, particularly the problem of deciphering accurate precipitation forecasts in the numerical weather prediction models\cite{reichstein2019deep}. It is important to note that the progress made by dynamic models should not be ignored in favour of deep learning, but rather should be supplemented by this new technique   \cite{dasgupta2020exploring, singh2020overview} . The study by Reichstein et al.  \cite{reichstein2019deep}  showed how deep learning could be applied to geoscience problems and pointed out that with the abundant growth of data and computing resources, machine learning advancements have yielded transformative results across various scientific disciplines.  They also indicated that machine learning is becoming a popular approach to transformation and anomaly detection, and geoscientific classification. The use of neural networks to detect extreme weather patterns replacing traditional threshold-based analysis is one example.

Deep learning is a data-hungry method that can learn the complex mapping between inputs and outputs. This method has shown remarkable results in various fields including meteorology, where it can be used to forecast the precipitation \cite{shi2017deep} . The efficacy of deep learning approaches in predicting precipitation is explored in this study. We have extensive meteorological data, such as ground-based and remote sensing based satellite observations, collected by various techniques for the past several years. One such important meteorological variable of interest to Indian community is precipitation. This study aims to develop a deep learning model for forecasting spatio-temporal sequences and apply it to ISM precipitation data. Vishwnath et al. \cite{viswanath2019deep} attempted to study the active and break spell of monsoon using Long Short Term Method (LSTM)-     based networks. The classification problem was approached with a LSTM based model and a seq-seq model, which contained a LSTM encoder-decoder with an attention mechanism. These LSTM based models were found to outperform the other machine learning-based models like Support Vector Machines (SVM) and K-Nearest Neighbors (KNN). Also, a weighted softmax function was implemented to counter the effect of imbalances in data. The study in \cite{siami2018comparison} also shows the power of LSTM which outperformed the Autoregressive Integrated Moving Average (ARIMA) model by reducing the error rate up to 87\%.

Further studies, such as \cite{khan2020hybrid} , have shown the effectiveness of using a hybrid model with conv2D and Multi-Layer Perceptron (MLP) to do a multivariate prediction for rainfall. When compared with a simple MLP and an SVM, this hybrid model was found better. The convolutional 1D and MLP together better captured the complex relationship of rainfall with the other variables.  The work by researchers in \cite{ham2019deep, saha2020prediction} demonstrates the power of convolutional neural-network-based architecture  to predict the El Nino–Southern Oscillation (ENSO) variations effectively. Their model was able to give skillful forecasts for lead time up to one and a half years. The nino3.4 index of the model was found to be better than other state      of the art dynamic models. Most of the above models used either only convolutional or LSTM based architectures to capture rainfall patterns. These models also only tried to either classify or detect patterns in the future.   For predicting  the rainfall values,  a model has to be more powerful, able to capture temporal and spatial structure of the data and hence, we note the usage of ConvLSTM based architectures in \cite{shi2017deep, kim2017deeprain,xingjian2015convolutional} .  In \cite{shi2017deep} the effectiveness of ConvLSTM over linear regression is established by working with multichannel radar data.  

\begin{figure*}[!t]
	\centering
	\includegraphics[scale=0.5]{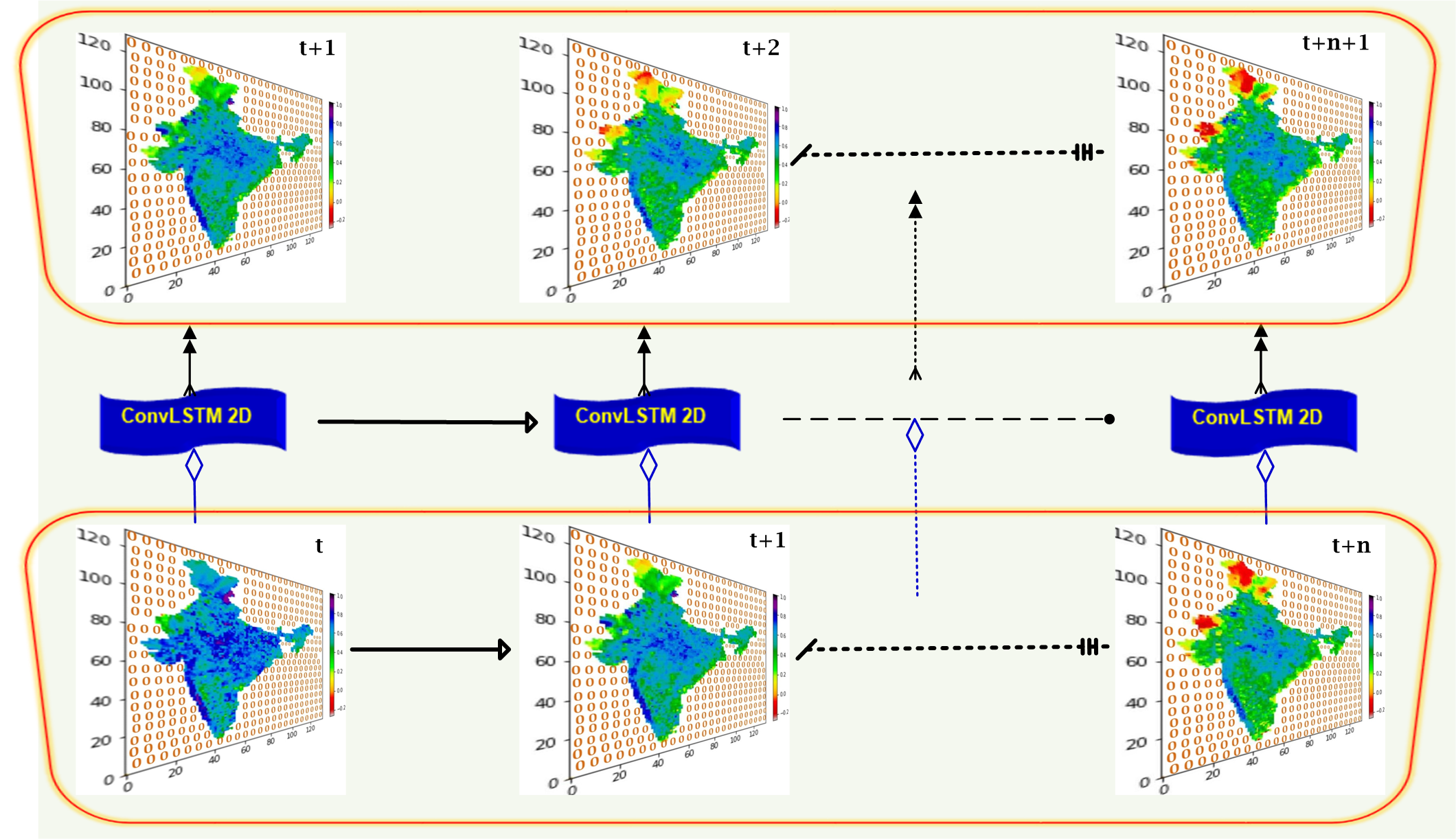}
\caption{ A sketch of ConvLSTM architecture based on (Shi et al., 2015)  used in this study.
}
\label{fig:cnv_intro}
\end{figure*}

ConvLSTM model is a hybrid model that uses the spatio-temporal information to generate the forecast.  For dispersive waves (such as Rossby waves, convectively coupled waves) which have typical wave-frequency spectral signatures and generate skewed weather states (e.g. extreme weather events), this type of spatio-temporal information based model is a natural choice. For the present study, we therefore chose this model as we want a state-of-art model \cite{xingjian2015convolutional}  which is already successful in similar applications but has not been applied for the monsoon forecast. There are not many monsoon forecast models available in literature which applies convolution and Recurrent Neural Network (RNN) based techniques. The  research  by Shi et al. in \cite{xingjian2015convolutional} is the best work for application of ConvLSTM, where the model's effectiveness is established for spatio-temporal sequence prediction problems. The ConvLSTM based model was also shown  to outperform the state-of-the-art optical flow-based ROVER algorithm  \cite{xingjian2015convolutional}. The above points motivate us to utilize it for precipitation forecasting in this study. A sketch of the network used in this work, based on \cite{xingjian2015convolutional}, is shown in Figure \ref{fig:cnv_intro}.

In this study, we have worked on two types of Geoscience data  for forecasting of precipitation. One of them is the ground-based in-situ precipitation data from the India Meteorological Department (IMD) and the other is remotely-sensed Tropical Rainfall Measuring Mission (TRMM) data which includes data from (i) Lightning Imaging Sensor (ii) TRMM Microwave Imager, and (iii) Visible Infrared Scanner. The next section provides details of the data and methodology used in this study. An efficient approach used for data pre-processing is described in Section III. The problem statement is described in Section IV and the architecture of Artificial Intelligence (AI) model developed for sub-district (25× 25km) scale and aimed towards short range (1-3 days) forecasting is explained in section V.  Section VI provides descriptions       of the results obtained from the model. This study's conclusions are contained in section VII.  The discussion and future work are provided in the last section.

\section{METHODOLOGY AND DATA}
The LSTM networks were first introduced by Hochrereiter and Schmidhuber  \cite{hochreiter1997long}. It typically has a forget gate, an input gate, an output gate with its weights in which it can control what information to retain and what to forget, thus learning long-term associations. Shi et al. \cite{xingjian2015convolutional} developed the architecture of ConvLSTM when designing a model for learning spatio-temporal correlation in precipitation nowcasting problem. In this architecture, convolutional operations replace the typical fully connected architecture within LSTM.

In a Fully Connected LSTM (FCLSTM), the inputs and outputs are 1D vectors transformed by weights through standard matrix multiplication. However, in a ConvLSTM cell, the standard matrix multiplications are replaced with weights performing convolution operations, as shown in the equations  (1a-1e) below \cite{xingjian2015convolutional}.

\begin{subequations}
\begin{equation}\label{conv_eqn}
i_t = \sigma( W_{xi} \ast\chi_t + W_{hi}\ast H_{t-1} + W_{ci} \circ C_{t-1} + b_i) 
\end{equation}
\begin{equation}
f_t = \sigma( W_{xf} \ast\chi_t + W_{hf}\ast H_{t-1} + W_{cf} \circ C_{t-1} + b_f)
\end{equation}
\begin{equation}
C_t =f_t \circ C_{t-1} + i_t \circ tanh (W_{xc} \ast X_t +W_{bc} \ast H_{t-1} +b_c)
\end{equation}
\begin{equation}
o_t = \sigma( W_{xo} \ast\chi_t + W_{ho}\ast H_{t-1} + W_{co} \circ C_{t-1} + b_o)
\end{equation}
\begin{equation}
\hspace{-5 cm} H_t = o_t \circ \,\, tanh(C_t) 
\end{equation}
\end{subequations}

In the above equations, `$\ast$’ stands for convolution operator, and `$\circ$’ stands for the Hadamard product (Elementwise matrix multiplication). The terms $i_t$ , $o_t$ and $f_t$ represent the input, output and forget gates, respectively. $W_{x\cdot}$ are weights calculated in different gates. The variables $\chi_t$ and $C_t$ are for the inputs (rainfall at day t) and outputs. $H_t$ denotes the hidden state, and variable represented by $b_{i,f,o}$ are the biased calculations in respective gates. All the gates and cell variables are the 3D tensors.

The most significant advantage when comparing ConvLSTM and FCLSTM is the reduction in the number of parameters in ConvLSTM. Shi et al. \cite{xingjian2015convolutional} listed the efficacy of ConvLSTM in capturing moving objects compared to FC-LSTM.   In this study, motivated by the work in \cite{xingjian2015convolutional} , we employed the ConvLSTM method for ISM Rainfall (ISMR) short range forecasting for up to 2 days lead time. 
\begin{figure}[!t]
	\centering
	\includegraphics[scale=0.40]{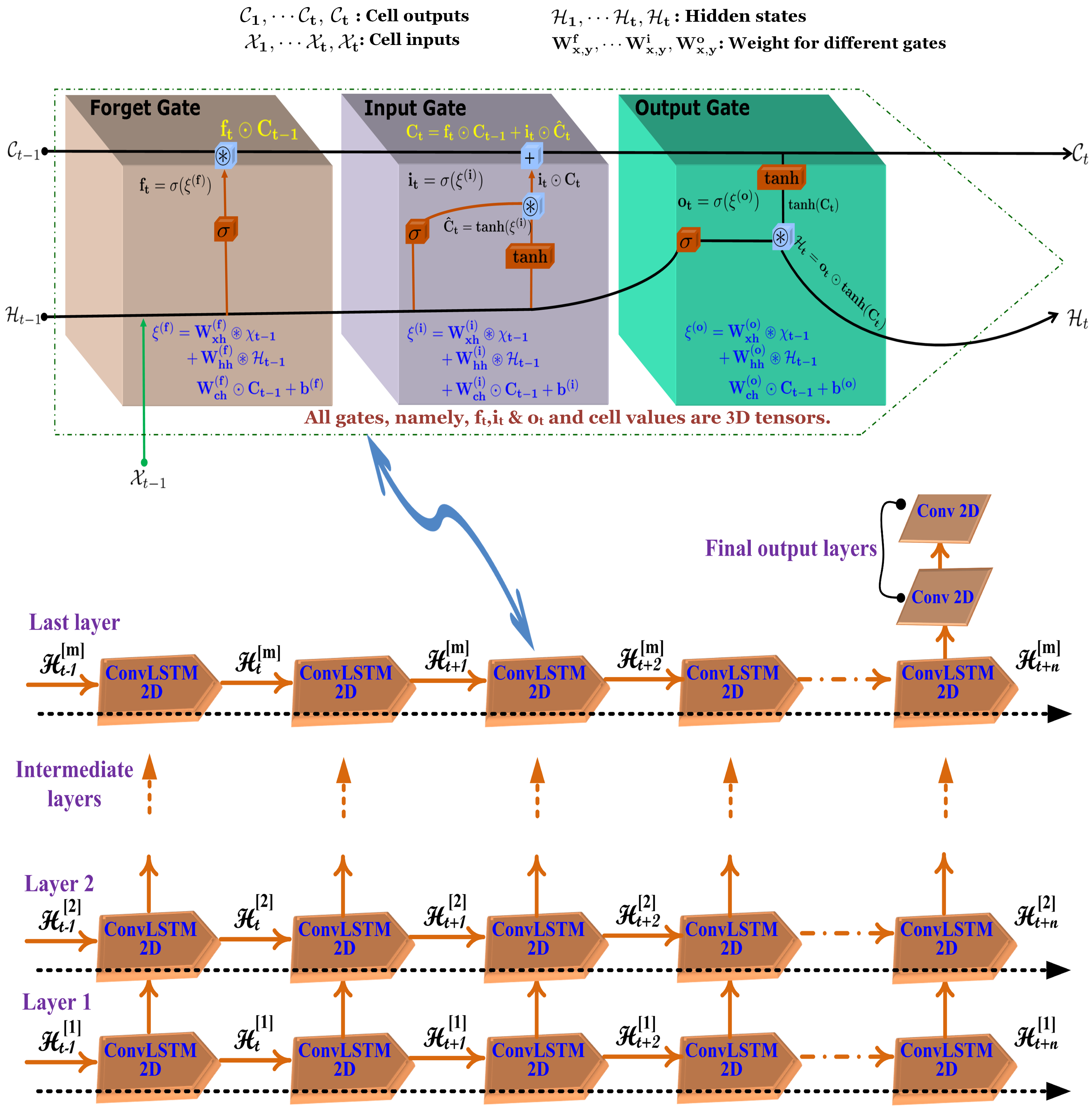}
	\caption{ The general stacked ConvLSTM architecture used for sequence forecast. The upper part of the figure provides details of a ConvLSTM 2D cell, and the lower part is for the model network used in this work. Three layers model was used for TRMM data, while in the IMD data model, five layers were used. The TRMM data did not require much pre-processing such as transformation from real to exponential space unlike IMD data hence, only three layers were sufficient. IMD data needed more efficient edge detection near the land-sea boundaries. Therefore, additional convolutional layers were added after experimentation. 
	}
	\label{fig:cnv_detail}
\end{figure}

An overview of this method is presented in Figure \ref{fig:cnv_detail}, explaining the entire model architecture used and details of a particular ConvLSTM cell. The lower part of Figure 2 illustrates a general architecture of a stacked network using the ConvLSTM cells.  The full details of a specific cell are provided in the upper part of the figure. Each cell contains 3 gates, namely, ‘Forget Gate’, ‘Input Gate’ and ‘Output Gate’. The functioning of these gates can be seen in \cite{xingjian2015convolutional}. 

For the data set used in this work, the activation functions ‘tanh’ and hard ‘relu’ for recurrent activation were found to be the best setting for ConvLSTM layers. Both activation functions were used in the developed model as shown in Table 2. In the model architecture (more details are provided in section V), the outputs can be taken from the network in two ways;      (i) sequentially, or (ii) only the output corresponding to the final time step (which inherently contains information on all previous time steps). A final layer of Conv3D or Conv2D (see lower panel of Figure 2) is added with a relevant activation function suited to the problem to form the final output. For this study, we applied two Conv2D layers for the final output in both data sets. More details of the internal layers of the architectures are provided in section IV, describing components of the model.

% needed in second column of first page if using \IEEEpubid
%\IEEEpubidadjcol

\subsection{Dataset}\label{sec:dataset}
The ConvLSTM model was tested on two main datasets. They are station based IMD gridded data \cite{pai2014development}  and remote sensing based TRMM data \cite{huffman2016trmm}.  Details of these datasets are given below:

The IMD dataset is obtained from interpolation of ground station data into a gridded form \cite{pai2014development} .  Rajeevan et al. \cite{rajeevan2006high}  studied the break and active spells of ISM using a high-resolution gridded dataset. The dataset was created from the ground data obtained from various stations across India. The stations were chosen based on their density to avoid any inhomogeneities. The Shepard interpolation, based on weights calculated from distance to nearest grid point and direction, was applied for generating the interpolated values. This effort   generated a ground-based daily gridded data with resolution of  $0.25^o \times 0.25^o$   over India, which was found to be more accurate than the other global gridded datasets \cite{rajeevan2006high}. We use this data for the period 1974-2015. 

NASA and Japanese Space Agency Jointly own the TRMM which contains the data obtained from satellite measurements and the same is available globally from 50o N to 50o S. The TRMM source data is in mm/hr unit, therefore a factor of 3 is multiplied to the sum for every grid cell. We have used the daily accumulated precipitation (mm/day) product for the period 1998-2015 from research quality 3-hour TRMM Multi-Satellite Precipitation Analysis (TMPA-3B42). The resolution of the data was  $0.25^o \times 0.25^o$  having invalid values which were set as -9999.  The TRMM accumulated precipitation is obtained as follows:
\begin{eqnarray}
P_{daily} = 3 \times \sum(P_i \times Valid (P_i)) \\
P_{dailycount} = \sum (Valid (P_i))
\end{eqnarray}

Where $Valid (P_i) =0$, if the data point is absent otherwise 1. The data set is available from January 1, 1998 to date. We have chosen the data for the present study till December 3, 2015 and between $6.375^o$ N to $38.625^o$ N and $66.375^o$ E to $100.125^o$ E.

Thus, both data were utilized on a daily basis, with each frame reflecting the total rainfall of the day. A sample of total rainfall for a particular day from these datasets is shown in the Figure 3.

\begin{figure}
	\centering 
	\includegraphics[scale=0.40]{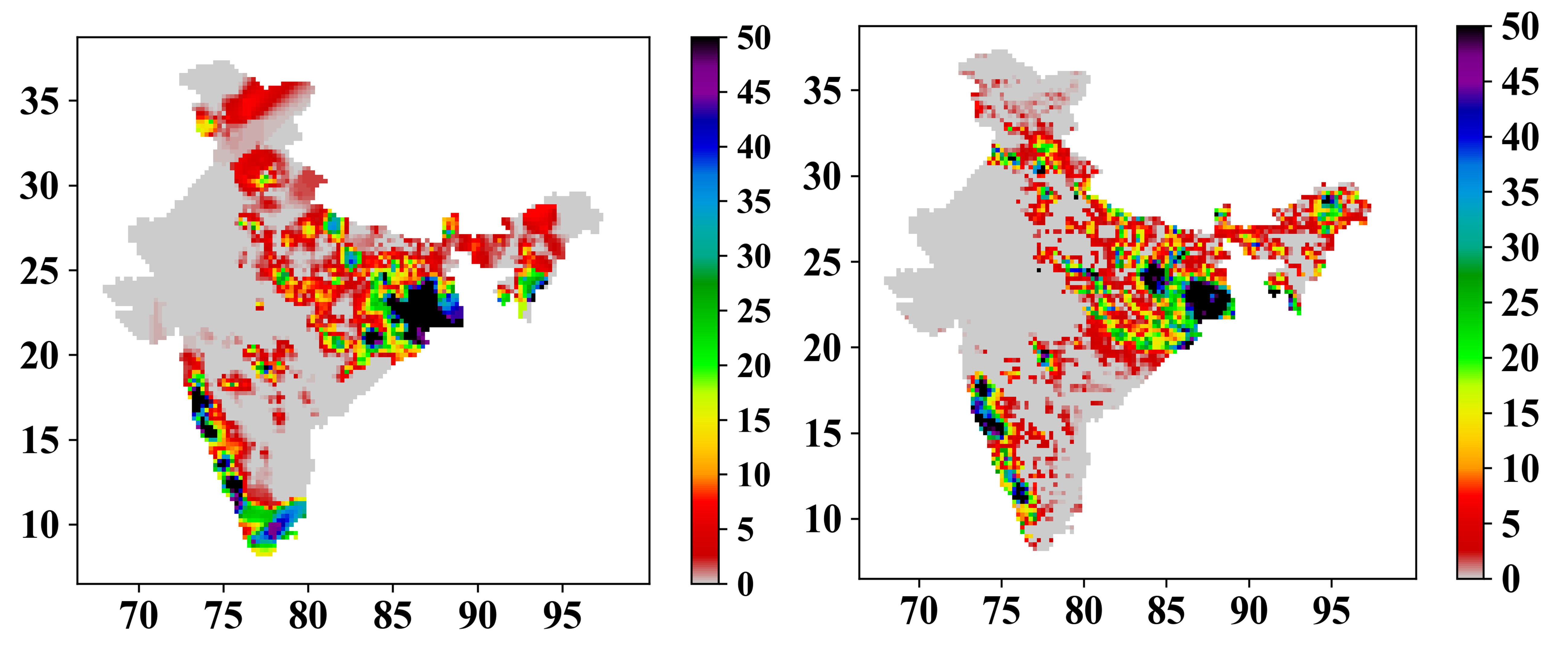}
	\caption{: Sample total rainfall on a day (18-06-2011) from the IMD high-resolution dataset (in the left panel). The right panel shows a typical rainfall on a day from TRMM high-resolution dataset.}
	\label{fig:sample_data}
\end{figure}

\subsection{Data processing}\label{sec:data_process}
The ConvLSTM network used in this work receives data in 5 dimensions, namely: no. of samples, time steps, latitude, longitude, and variables. It is essential to clean up and prepare the data in a supervised learning format. During the processing of data for both datasets, different techniques described in the following subsections have been adopted.

\subsubsection{Station based (IMD) Dataset}\label{sec:IMD_data}
This dataset had several undefined values which were assigned as `NaN'. There were some points which were assigned as `NaN' in all frames and others were those which were rarely absent. The points under the second category were interpolated spatially from their closest neighbors.  Special treatment had to be given to the points having `NaN' in all frames to avoid losing spatial structure of the data while treating NaN values. We have implemented a new and efficient method for this problem, detailed in Section III. 

\subsubsection{Remote sensing based (TRMM) Dataset}\label{sec:TRMM_data}
There are a significant number of invalid points within the TRMM Dataset. We spatially interpolated them from the nearest neighbours. Although a high degree of skewness was a specific difficulty that was faced while dealing with this data, we wrote a custom loss function for TRMM data training. A similar approach was used in \cite{shi2017deep}. The details of the custom loss function are provided in subsection C1.

\subsection{Metrics for assessing the robustness of results}\label{sec:Metric}
 We have used a new custom loss function ($\lambda_{mse}$) to deal with the invalid points in the TRMM dataset. It is defined in equation (4).  To validate our model we used correlation coefficient (CC),calculated based on predicted and true values as given in equation 5. Furthermore, we calculated the ROC curves (using equation (7)) to analyze the skill of the forecasts. The details of these metrics are provided in the following subsection.
 
 \subsubsection{A New custom loss function for TRMM training}\label{sec:custom_loss}
\begin{figure}
	\centering 
	\includegraphics[scale=0.40]{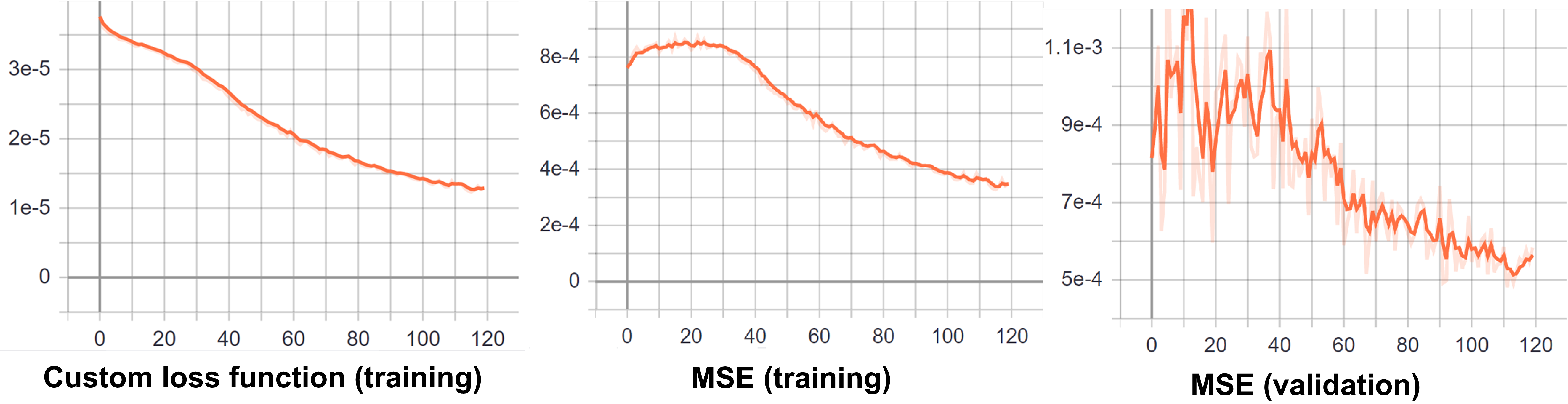}
	\caption{Comparing the variation of custom loss with MSE on TRMM Data (X-axis- epochs, Y-axis- Error).}
	\label{fig:MSE_comparison}
\end{figure}

Since the TRMM data training is a regression problem, Mean Squared Error (MSE) is the usual choice of the objective function (loss function). However, the skewness in the data resulted in the model not predicting large values in the ground truth ($>30mm$). Therefore, the model was trained with a custom loss function ($ \lambda_{mse}$) given as follows: 
\begin{subequations}
\begin{equation}\label{eqn:loss_function}
\lambda_{mse}=\frac{1}{N}\sum_{1}^{N}\sum_{1}^{N_{lat}} \sum_{1}^{N_{lon}} W_{n,i,j} \ast (x_{n,i,j} - \tilde{x}_{n,i,j})^2
\end{equation}
\begin{equation}\label{eqn:w1}
\hspace{-3.5 cm}  W = \ 1 \ \ \  if \ \  x_{i,j}>= 0.15
\end{equation}
\begin{equation}\label{eqn:w0}
\hspace{-3.5 cm}  W =0.1 \ \ if \ \  x_{i,j} \ <  \ 0.15
\end{equation}
\end{subequations}
Here $x_{ij}$ represents the value from the normalized TRMM dataset. The choices for these hyper parameters in equation  \ref{eqn:w1} and \ref{eqn:w0} were arrived at by trial and error approach. A higher weightage needed to be given to the higher value because the rainfall data was skewed and the extreme events were required to be captured appropriately. The choices of the limits and the weight are empirical.  A comparison of the custom loss function with MSE defined in equation (\ref{eqn:mse}), is presented in Figure \ref{fig:MSE_comparison}. In the figure X-axis and Y-axis represent epochs and error respectively. Training was stopped at early stage as shown because no further reduction in validation loss was found after those epochs.

\begin{eqnarray}\label{eqn:mse}
\hspace{-0.5 cm} MSE = \frac{ \sum_{ N_{samples}} (\sum_{ N_{lat}} \sum_{ N_{lon}} ( y_{pred}-y_{true})^2 ) } {
N_{samples}N_{lat}N_{lan}  }
\end{eqnarray}

\subsubsection{Correlation}\label{sec:correlation}
In meteorology and geophysical fields, we generally get the data in the 3-dimensional space in particular, any variable in the data can be represented as $x \in [L_1,L_2,T]$, where L1 is latitude, L2 is longitude and T represents time. That means the data is in the form of space coordinates which represents spatial pattern maps in  $[L_(1,) L_2 ]$ plane for a given time slice. One can have a temporal correlation between two variables at a given location, for a set of time coordinates, or alternately, for a given time, a correlation between the two variables for spatial locations. This metric is known as the pattern   correlation coefficient (CC). It signifies that for a given time how the spatial variances are related between two variables. In other words, it represents how well the two variables (say rainfall from observation and from forecast) are spatially collocated.  It is calculated with the following formula \cite{weisstein2020statistical}.

\begin{equation}
CC =  \frac{ \sum ((y_{pred} - \mu_{pred}) (y_{pred}- \mu_{pred}) ) } { \sqrt{ \sum (y_{pred}- \mu_{pred})^2  \sum (y_{true}- \mu|{true})^2 } }
\end{equation}

Here, the summation is taken over the test data. We calculated this metric for  TRMM data set, with corresponding IMD data, shown  in the results section.  Apart from the custom loss function for the TRMM data, we have used an efficient approach for dealing with `NaN' values in IMD data, described in the next section. 

\subsubsection{Receiver Operating Characteristics (ROC) curve}
Another metric, used in this work, to validate the results is ROC. It is an important tool for forecast verification and decision-making processes.   It is a plot which illustrates the diagnostic ability of an forecast classier system, using its varying discrimination threshold (see \cite{marzban2004roc} ).  The ROC curve is created by plotting the hit rate or True Positive Rate (TPR) against the False Positive Rate (FPR) at various threshold settings.  The ROC analysis provides the ways to select possibly optimal models and to discard suboptimal ones independently from (and prior to specifying) the cost context or the class distribution. This analysis is related, in a direct and natural way, to cost-benefit analysis of diagnostic decision making. Hence, it is a standard method of forecast skill analysis for operational rainfall forecast. While the correlation method can't discriminate the threshold criteria for more false positive occurrence, the ROC method can do so. The ROC method applied here shows a better fidelity of the proposed model. 

The formula for calculating these rates are given in equation (7). 

\begin{subequations}
	\begin{equation}\label{eqn:tpr}
	TPR =  \frac{TP}{N_H}  
	\end{equation}
	\begin{equation}\label{eqn:fpr}
	FPR = \frac{FP}{N_L}
	\end{equation}	
\end{subequations}

Where, TP denotes True Positive and it is number of days when both area averaged values of ground truth and prediction are above average. $N_H$ is the number of days when area averaged ground truth values is higher than the threshold values chosen based on minimum and maximum rainfall values in the data. FP denotes False positive and it represents days for area averaged value of prediction above the threshold when the prediction value is below the level. The number of days when area averaged ground truth rainfall values lower than the threshold is represented by $N_L$ in the equation \ref{eqn:fpr}. 

\section{METHOD FOR DEALING WITH ‘NAN’ VALUES}
To deal with `NaN' values in the IMD data a new strategy was employed in this work. A detailed description of the strategy is provided in this subsection. 

As mentioned in section \ref{sec:IMD_data}, two kinds of `NaN’ values were present in the data: (i) points which are `NaN’ in all frames which refers  to those points which correspond to ocean and lie outside India and, (ii) points that are occasionally missing due to lack of observation on a day because of equipment malfunctions etc. The occasionally missing `NaN’ points were spatially interpolated from their nearest neighbours values. 

The `NaN' values (in point (i) above) cannot be extrapolated as there is no sufficient data for so many points. Also they can’t be replaced by `0' because `0' number has a significant value for precipitation as it indicates no rains (depicted in Figure \ref{fig:real_exp}). It gives impression that there is no rain over the ocean regions which is wrong.  Therefore, in the real space, data would furnish wrong information to the model.  For such cases we need to use `0' values for training the model smoothly. One solution is to take the points into a 1D vector for each time step and then try out the mapping. However, this method of reducing the dimensions would destroy the spatial structure of the data and derail the whole purpose of using a ConvLSTM based architecture. Therefore, a new method was tried out in this work to deal with `NaN' values falling outside of the Indian landmass. This involves taking the data into exponential space and then assigning `0' for missing values represented by `NaN'. This is done keeping in mind the practice that; in general, it is safe to input missing points with ‘0’ provided that it doesn’t represent a meaningful value. The condition of `0' not being a meaningful value is not met in the real space because the locations with no rainfall are marked as `0' in the raw data. Therefore, an efficient transformation was required which we chose as exponential space as discussed in the previous paragraph and illustrated in Figure \ref{fig:real_exp}.

\begin{figure*}[!t]
	\centering 
	\includegraphics[height=0.40\textheight]{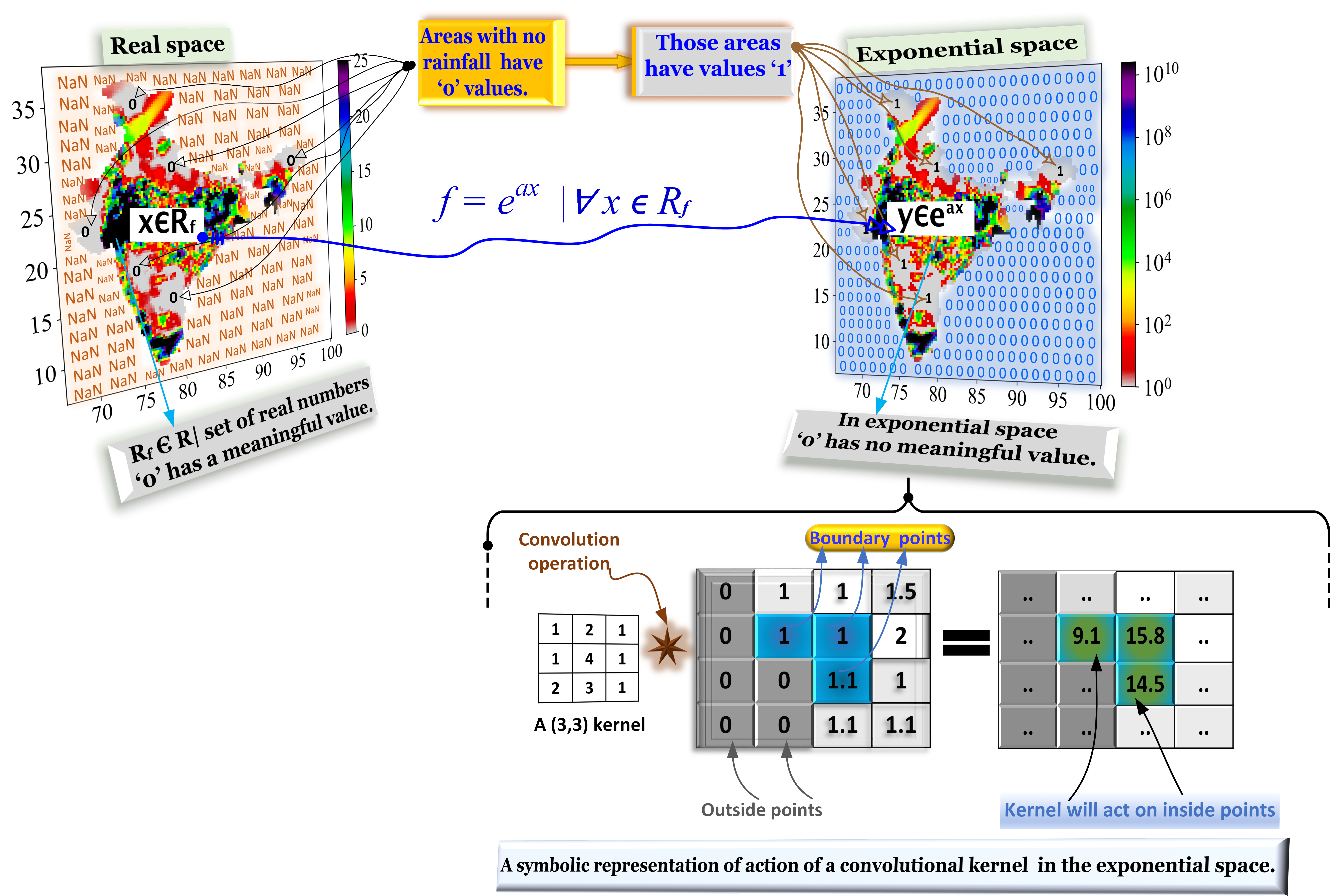}
	\caption{Illustration of data transformation from real space to exponential space. The conversion in exponential space has the benefit of considering ‘0’ during model training as it doesn't possess any meaningful value. }
	\label{fig:real_exp}
\end{figure*}

While converting the whole data from `real space' to `exponential space', we got rid of the issue of wrong information (Figure \ref{fig:real_exp}. The network learns from exposure to the data to treat the value `0' as missing and start  ignoring  them in the transformed space \cite{chollet2017deep} . We found that this method is best suitable for the model training in the present scenario since it is one of the effective techniques which can take care of sharp gradients in spatial patterns of rainfall that we see along the west coast of India. In meteorological data, dealing with the missing values is an essential problem and the said transformation is one of the effective methods which can be used operationally. The data preparation is done, as explained below (also refer to Figure \ref{fig:real_exp}).
\begin{itemize}
	\item[1.] Identify all non-NaN points and normalize them with maximum value in the dataset (for all points).
	\item[2.] Apply the transformation $f= e^{ax} \forall x \in  [ not-NaN]$.
	\item[3.] Apply zero to NaN in the rearranged dataset (legitimate values now range from 1 to $e^a$).
	\item[4.]The choice of `a' is appropriate when the range of the initial dataset and transformed dataset approximately match. In this way `0' rainfall value is transformed to 1 so the whole range of allowed values becomes $[1,\infty)$. 
	\item[5.] Network maps input to output in exponential space.
	\item[6.] The spatial structure of data is preserved; hence spatial correlations can be learned.
\end{itemize}
In our knowledge, no models in literature describe treating such missing values (i.e., where observation values are unavailable) in an effective way and it is the first time such a transform has been used in the field of meteorology for AI model training.  Hence, this method may be treated as a novel approach to deal with missing data values. 

\section{PROBLEM FORMULATION}
Usually, weather predictions come with probabilistic scoring, which is why problem statements of weather prediction can be written as most likely N-sequence selection from an ensemble of prediction. One such probability scoring is Continuous Ranked Probability Score (CRPS) as described in \cite{zamo2018estimation}.  But, for deterministic forecasts like neural networks, CRPS reduces to mean absolute error. As a spatiotemporal sequence forecasting problem (for monsoon rainfall), our input state can be represented as vectors of variables over a spatial grid of $L_1\times L_2$ locations as described in Section \ref{sec:correlation}.

On these locations, say, total $N_p$ variables are measured. Therefore, any observation at a given time is represented in a mathematical space $R^{(L_1\times L_2 \times N_p )}$, where $R$ is the domain of the observed variables. Given a certain periodicity of the past data, it can be represented as a sequence of elements from this aforementioned space as $X_1,X_2,X_3,\cdots X_t$. Then the forecasting problem is defined as to predict the least error K-length sequence in the future given the previous ‘t’ observations (including the current one) as input. This can be represented as

\begin{equation}\label{eqn:problem_formulation}
Y_{t+1}, \cdots, Y_{t+1} = f (X_{t+1}, \cdots, X_{t+k} | X+1,X_2,X_3, \cdots, X_t )
\end{equation}
where Y is the predicted output sequence and f  is the least-error forecasting function. The function f, here, is a high-dimensional parameterized function based on artificial neural network architectures. In other words, our problem reduces to finding a suitable architecture among various possibilities of hyper-parameters and layer choices which reduces the error between the predicted and the ground truth of observations. We started with simple ConvLSTM-based architecture and tuned it to improve our predictions but were constrained by the number of layers and layer-specific hyper-parameters that could be chosen given an upper limit of RAM and processing power of the Graphics Processing Unit (GPU).  As with any deep learning problem, some of these parameters have been improved but optimality cannot be guaranteed, in present scenario, for all hyper-parameters.

\section{THE AI MODEL ARCHITECTURE}
As mentioned before, we decided to employ the ConvLSTM method, thus developed the model for this algorithm and carried out several experiments to mature the architecture. The experiments were mainly based on data pre-processing and techniques used for handling the undefined rainfall values assigned as `NaN'. In the case of IMD data, we used the exponential space to train the model. 
Once the algorithm and kind of Neural Network architectures are decided, the network's fine-tuning, called hyperparameter optimization, is accomplished. Various combinations of kernel sizes, number of filters, activations, number of layers, optimization algorithm, and learning rate are tried out during training before asserting the best final architecture. 
For both datasets, the developed models were trained using the Keras API with TensorFlow running as a backend. The choice of the last layer to be fitted to the ConvLSTM output was selected from the following options:

\begin{itemize}
\item[1.] Conv3D Layer: This layer is applied to the 5-dimensional sequential output of the connected ConvLSTM layers. It performs a 3D convolution over space and time dimensions to produce the final output.

\item [2.] Conv2D Layer: To apply this layer, the ConvLSTM is set to return only the output corresponding to the last time-step in an input. Therefore, this layer uses a 2D spatial convolution on the spatial dimensions alone to give the output. 
\item[3.] Locally Connected 2D Layer: This layer acts similar to Conv2D but in a generalized form. The kernel used is different at each location throughout an image. It has more parameters compared to Conv2D, but spatially localized patterns could be learned.
\end{itemize}

The developed model used the Conv2D as the last layer based on the MSE value. A comparison of MSE among different layers is provided in Table \ref{tab:mse_layers}.

\begin{table}[ht]
	\caption{A comparison of MSE among different layers used as final layer. The least value was obtained using Conv2D layer, hence, it was chosen as last layer.}
	\label{tab:mse_layers}
\begin{tabular}[ht]{|c|c|c|c|}
	\hline
	Layer  & Conv3D  & Locally connected 2D & Conv2D \\
	\hline
	MSE    & $3.1\times 10^{-2} $  & $2.94 \times 10^{-2}$ & $2.76 \times 10^{-2}$ \\
	\hline
\end{tabular}
\end{table}

This study is an attempt to provide a proof of concept for applying the ConvLSTM method for ISMR forecasting. The study by Shi et al. \cite{xingjian2015convolutional}  proved that this method is better than other state of art machine learning methods available for forecasting meteorological variables. 
The details of the model architecture used for IMD and TRMM data are summarised in Table 2 and Table 3. The total number of parameters trained for IMD and TRMM datasets are 43559 and 284409, respectively. 

\begin{table}[ht]
 \caption{The model architecture used for training on the IMD rainfall dataset. Total 7 layers were used for this model.}
 \label{tab:model_architecture}
%\begin{tabular}{|c|c|c|c|c|c|}
%\begin{tabular}{|p{0.02\textwidth}|p{0.3\textwidth}|m{0.3\textwidth}|m{0.1\textwidth}|m{0.15\textwidth}|m{0.1\textwidth}|}
 % \begin{tabularx}{\columnwidth}{|c|c|c|c|c|c|}
 \begin{tabular}{|m{1em}|m{2cm}|m{2cm}|p{1 cm}|m{0.7 cm}|m{0.5cm}|}
  \hline
  LN  &   Layer name   & Architecture type & Activation & Kernel size &  \# Filter \\
  \hline
  1  & ConvLSTM2\_1  & Convolutional  LSTM & tanh & (3,3) & 4 \\
  \hline
   2  & ConvLSTM2\_2  & Convolutional  LSTM & tanh & (3,3) & 8 \\
  \hline
    3  & ConvLSTM2\_3  & Convolutional  LSTM & tanh & (3,3) & 8 \\
  \hline
    4  & ConvLSTM2\_4  & Convolutional  LSTM & tanh & (3,3) & 16 \\
  \hline
  5  & ConvLSTM2\_5  & Convolutional  LSTM & tanh & (3,3) & 16 \\
\hline  
  6  & Conv2D\_1  & Convolutional & relu & (3,3) & 15 \\
   \hline
  7  & Conv2D\_2  & Convolutional & relu & (3,3) & 1 \\
  \hline
 \end{tabular}
%\end{tabularx}
\end{table}

\begin{figure}[!t]
	\centering 
	\includegraphics[scale=0.4]{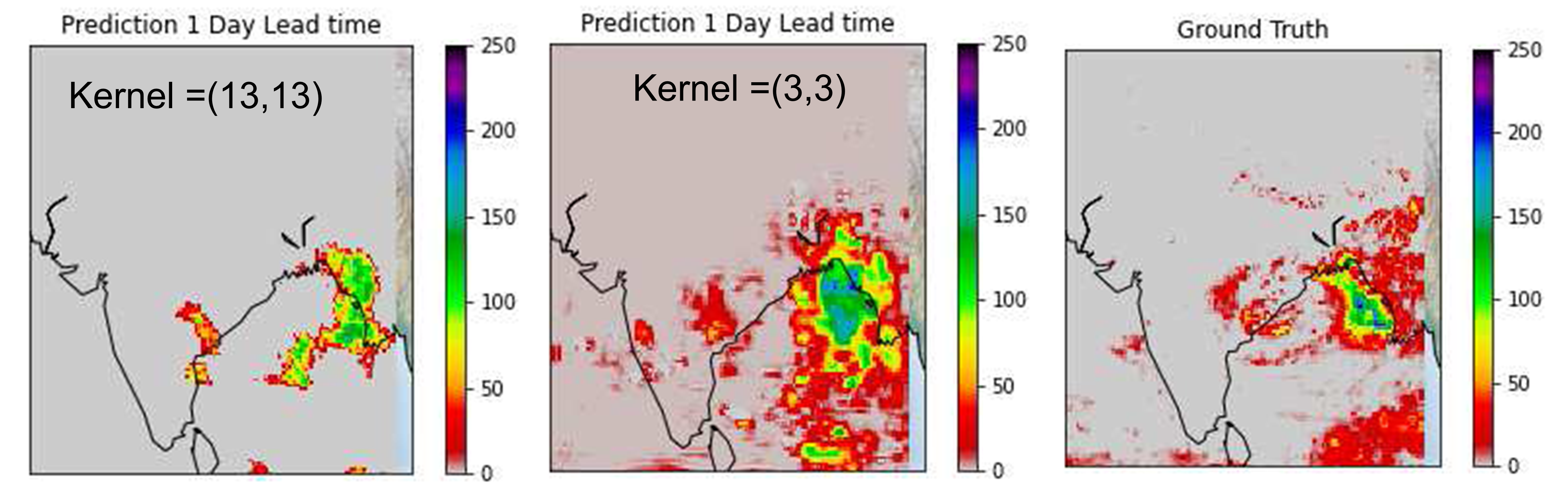}
	\caption{Comparing the 1 day lead predictions from a model using kernel sizes (13,13) and (3,3) with Ground truth, respectively (on TRMM Data). }
	\label{fig:real_exp}
\end{figure}

\subsection{Kernel size optimization }
Furthermore, we did experiments with different kernel sizes. It was observed that smaller kernel sizes tend to do better than larger ones. An example for 1 day lead time prediction is shown in Figure 6 when the Kernel is large (13, 13) and one with small (3, 3).  Figure 6 suggests that a smaller kernel of size (3, 3) can capture larger values effectively and also over more regions as compared to the larger one (13, 13). 

\begin{table}[ht]
	\caption{The model architecture used for training on the IMD rainfall dataset. Total 7 layers were used for this model.}
	\label{tab:model_architecture_trmm}
%\begin{tabular}{|c|c|c|c|c|c|}
\begin{tabular}{|m{1em}|m{1.8cm}|m{1.8cm}|m{0.9 cm}|m{0.6 cm}|m{0.6cm}|}
		\hline
		LN  &   Layer name   & Architecture type & Activation & Kernel size &  \# Filter \\
		\hline
		1  & ConvLSTM2\_1  & Convolutional  LSTM & tanh & (3,3) & 8 \\
		\hline
		2  & ConvLSTM2\_2  & Convolutional  LSTM & tanh & (3,3) & 12 \\
		\hline
		3  & ConvLSTM2\_3  & Convolutional  LSTM & tanh & (3,3) & 6 \\
		\hline
 		4 & Conv2D\_1  & Convolutional & relu & (3,3) & 6 \\
		\hline
		5  & Conv2D\_2  & Convolutional & relu & (3,3) & 1 \\
		\hline
	\end{tabular}	
\end{table}

\subsection{Computational resources}

The training was done at Pratyush HPC at the Indian
Institute of Tropical Meteorology, Pune \cite{pratyush_hpc}. The Pratyush HPC has a separate research and development cluster known as the XC-50 system. This system has 16 accelerator nodes powered with 1 Tesla P100 GPU (12GB RAM) and 1 Intel Xeon(R) CPU having 64 GB RAM.  

\section{EXPERIMENTAL RESULTS }

We solved a regression problem rather than classification as described in equation (\ref{eqn:problem_formulation}) in the section IV. However, classifications are made to understand the fidelity of the generated forecast. Normally, it is known that forecasts are skillful for rainfall above or below certain amplitude (or certain frequency). Verification of meteorological forecast is made in multi-category classification to emphasize the more skilful category. Operational forecasters always require such information to see the reliability of the forecast when the output values are above a certain threshold. The categories are made based on standard World Meteorological Organization manuals. As mentioned in the section II, we considered three different metrics to verify our forecast. The results obtained from model and analysis of metrics are presented in this section.

We applied the ConvLSTM algorithm on two sets of data: IMD gridded data and TRMM satellite data.  
Since both data sets have different pre-processing, as discussed in section II, two separate models were developed for them and were tested on validation data as given in Table \ref{tab:dataset}.

\begin{table}[ht]
	\caption{Details of the data segregation for training and testing purposes.}
	\label{tab:dataset}
	\centering
	\begin{tabular}{|c|c|c|}
	\hline
	 Data set  & Training set  &	Testing set \\
	 \hline
	 IMD	   &  22 years	   &     8 years  \\
	 \hline
	 TRMM	  &   12 years	   &      6 years \\
	 \hline
	\end{tabular}	
\end{table}

Out of the available 30 years IMD data, 22 years data was chosen for training and the remaining 8 for testing. For the TRMM dataset, 12 years were set for training while the remaining 6 for testing. The resulting outputs were compared with ground truths      using       metrics including correlation.  

\subsection{Comparison with Ground Truth}
The outputs obtained by applying models on both data sets were compared with available ground truth.  Details of these comparisons are provided in this subsection.

\subsubsection{IMD homogeneous regions data:}
\begin{figure}[ht]
	\centering 
	\includegraphics[scale=0.46]{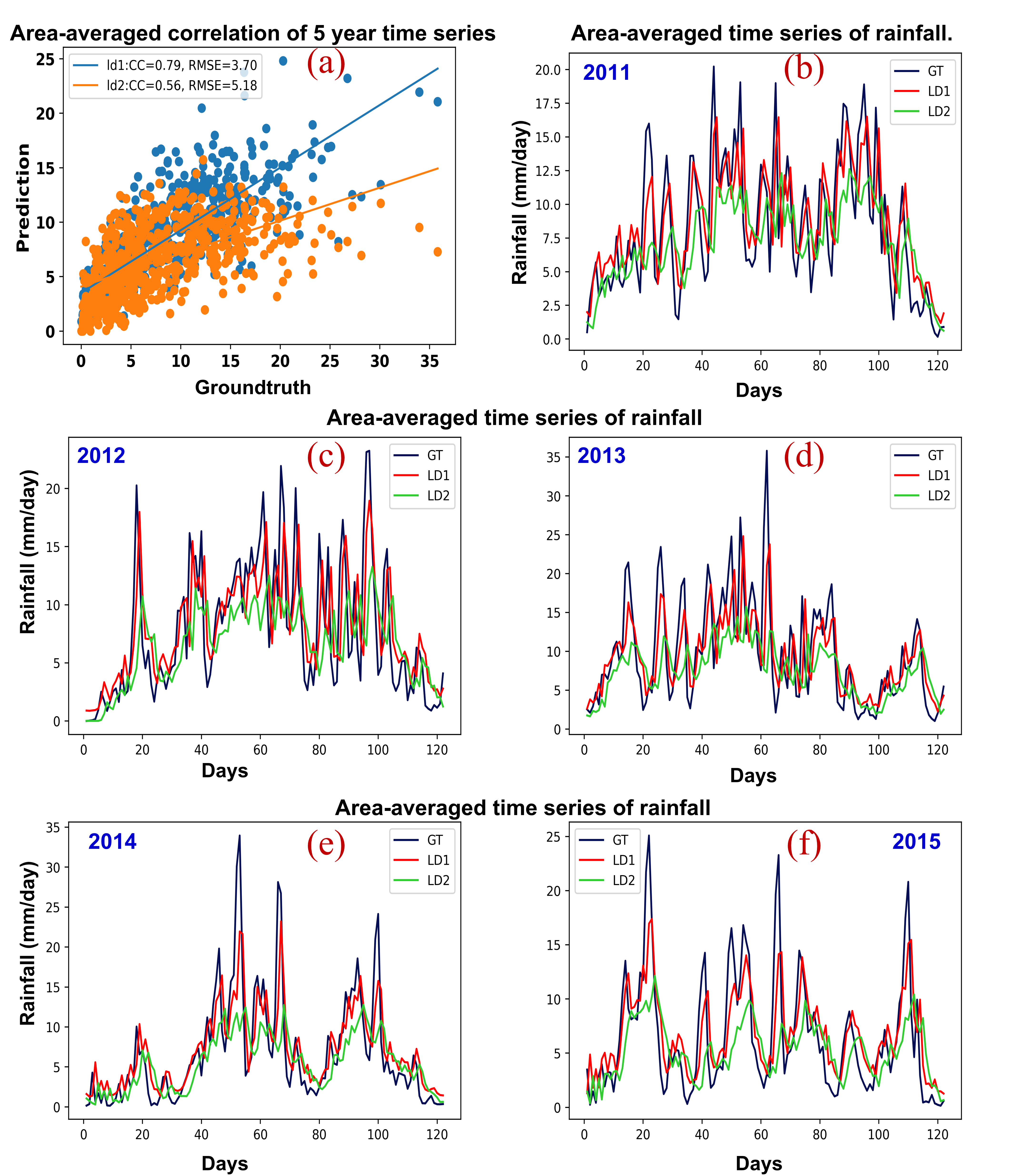}
	\caption{:  Comparison of area-averaged correlation coefficient (CC) and RMSE (panel a) for 5 years time series data and area-averaged rainfall for 5 years duration (2011-2015) for West Central  region (panel b-f). The days on X- axis starts from 1 June. The ld1 refers to lead day one and similarly ld2. }
	\label{fig:average_CC_west}
\end{figure}

First, we analyzed predicted data from the model for the homogeneous regions defined by the Indian Meteorology Department (IMD) \cite{kothawale2017monthly}. There are a total of 6 homogeneous rainfall regions categorized based on the rainfall percentage in monsoon seasons during the period from 1871-2016.  We calculated the Coefficient of Correlation (CC) for area-averaged data for 5 years’ time series and area-averaged rainfall for the 5 years duration from 2011-2015 for the IMD homogeneous regions. A comparison of these metrics with ground truth and model data for the west-central region is shown in figure  \ref{fig:average_CC_west}.

\begin{table}[ht]
	\caption{List of skills metric for different homogeneous regions. The correlation coefficient (CC) drops from 0.79 (West Central) to 0.52 (South Peninsular).  There is no specific trend for RMSE values.}
	\label{tab:skills}
	\centering
	\begin{tabular}{|c|c|c|}
		\hline
		West Central  &	0.79 (3.70)   &	0.56 (5.18) \\
		\hline
		Central NE	  &  0.7 (3.92)	  & 0.42 (5.11)  \\
		\hline
		Northwest     &	0.76 (3.77)	  & 0.58 (4.64)  \\
		\hline
		Hilly Regions &	0.53 (4.19)	  & 0.24 (4.93)  \\
		\hline 
		Northeast     &	0.55 (5.85)   &	0.3 (6.84)  \\
		\hline
		South Peninsular &	0.52 (4.15) &	0.31 (4.46)\\
		\hline
		
	\end{tabular}	
\end{table}

\begin{figure}[!t]
	\centering 
	\includegraphics[scale=0.46]{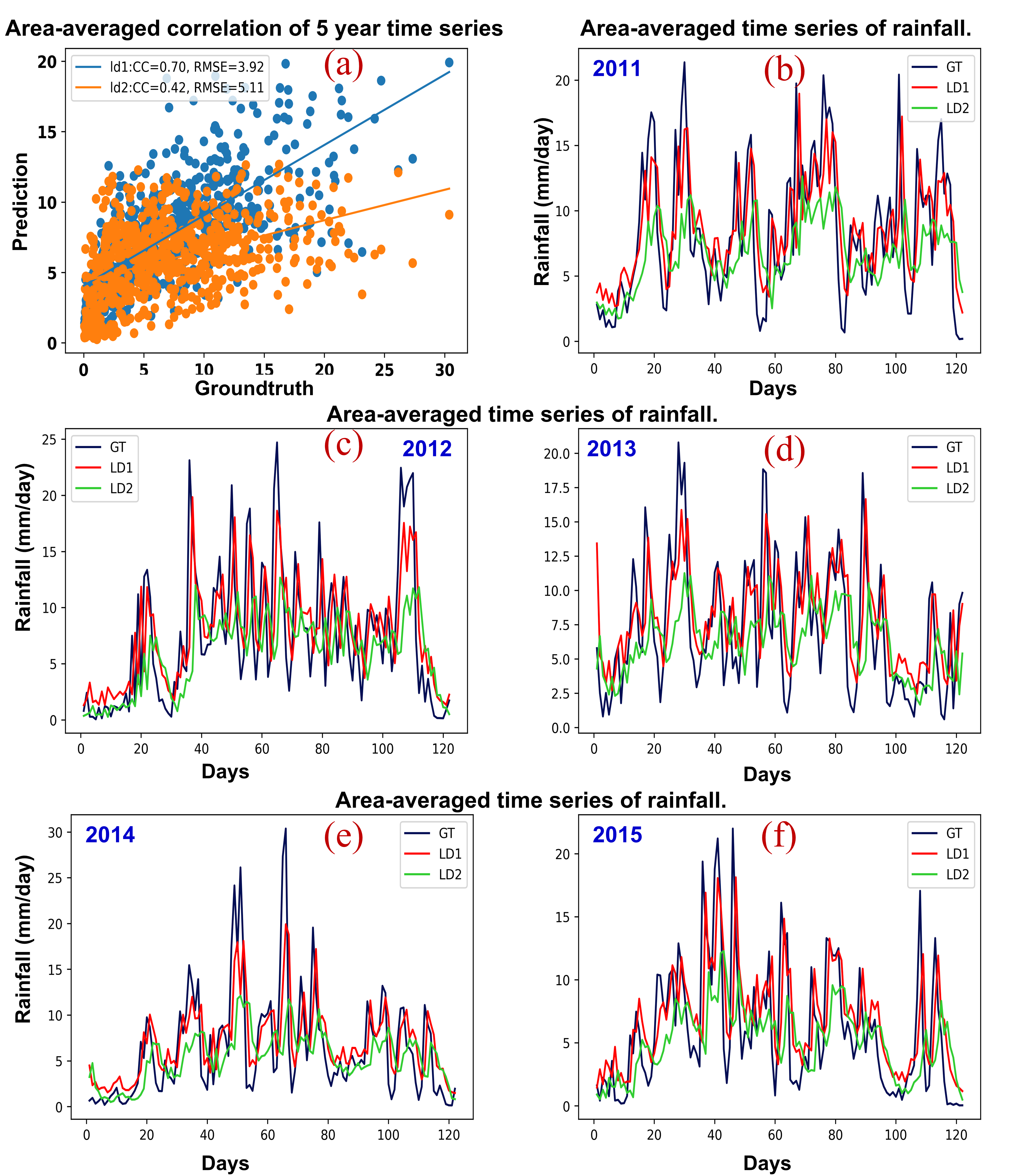}
	\caption{:  Comparison of area-averaged correlation coefficient (CC) and RMSE (panel a) for 5 years  time series data and area-averaged rainfall for 5 years duration (2011-2015) for Central North East region (panel b-f). The days on X- axis starts from 1 June. The The ld1 refers to lead day one and similarly ld2. }
	\label{fig:average_CC_central_NE}
\end{figure}

It is to be noted that the model can capture the rainfall up to 2 days lead time in the central region. A similar comparison for the Central North East region is provided in figure \ref{fig:average_CC_central_NE}. The skills for other homogeneous regions are presented in the Table 5. The CC values in this table  varies between 0.79 over West Central region to 0.52 over South peninsular. The CC and Root Mean Square Error (RMSE) values, obtained from this model, are comparable to state-of-the-art dynamical models such as present as shown by \cite{mukhopadhyay2019performance}.

\subsubsection{Comparison using entire data}

\begin{figure*}[ht]
	\centering 
	\includegraphics[scale=0.46]{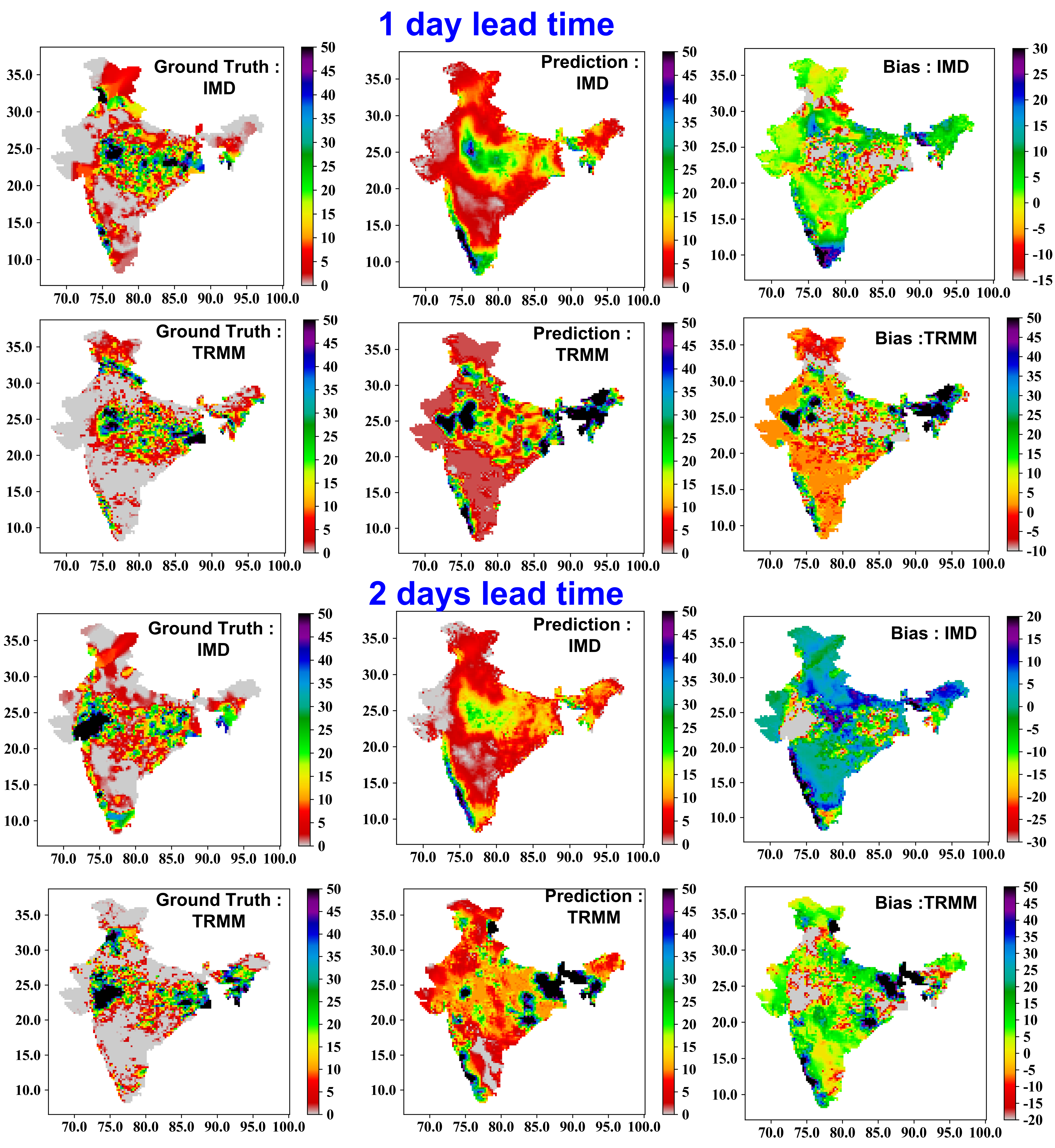}
	\caption{:  Comparing the 2 days lead predictions from the ground truth. The  upper panels to show the comparison for IMD data, and the  lower ones are for TRMM data in both ld1 and ld2 cases. The plot for ld1 is for August 8, 2011 (IMD), and August 7, 2011, for TRMM. The plots in the last columns represent biases. A similar comparison was present output obtained from dynamical model in \cite{huffman2016trmm}. }
	\label{fig:forecast}
\end{figure*}

Calculating the area average rainfall and comparing it with the ground truth for the homogeneous region is one way to test the model's accuracy. Further, we compared the spatial pattern of the forecast skill of the precipitation forecast for up to 2 days lead time for IMD and TRMM data for every grid point. One such comparison is depicted in Figure \ref{fig:forecast}. The TRMM dataset can capture localized as well as large-scale organized precipitation patterns. Previous studies have noted the capability of TRMM derived precipitation in capturing rainy spells and the extremes. It is beneficial over the regions of complex topography where in-situ data are often not available. However, it also predicts some false positives, predicting rainfall at places, not in the ground truth. The data was taken for August 8, 2011, for IMD, and August 7, 2011, for TRMM. The difference of 1 day between TRMM and IMD is due to the convention that IMD rainfall for a day is the rainfall obtained in the last 24 hours of the recorded time, while for TRMM, it is the rainfall in the next 24 hours of the recorded time.

The ISM rainfall shows significant variability in space  and time. On some occasions when the monsoon is in ‘active or organized’ phase, the rainfall patterns are widespread in space while during the ‘break or weak’ phase we see isolated spells across the region \cite{singh2021linkage} . It is to be noted that the rainfall memory (in time) is less as compared to other meteorological variables (e.g. temperature). Our aim here is to understand how well the model retains this memory and produces rainfall in space and time. Figure 9 compares the 1 and 2 day lead predictions generated by the model with the IMD and the TRMM data. It is to be noted that the training of the model was performed for both sets of data (the training periods were different). Therefore while comparing the model forecasts, corresponding observations are also used. The observation days here correspond to the model lead days and the bias is nothing but the difference (in space) between the observed rainfall for that day and the corresponding model forecast. It is seen that overall biases in both first (denoted ld1) and second day (denoted as ld2) lead times are smaller for the IMD data compared to the TRMM data. Though rainfall over the core monsoon zone shows less bias, there is significant bias over the regions of high elevations for both cases (e.g. over the Western Ghats, the Himalayan region). The analysis presented here helps us to identify the regions  where the model has good or poor fidelity in reproducing the actual rainfall and also indicates the spatial coherency between the two.

\subsection{Calculation of pattern correlation}
We calculated the pattern correlation as described in section \ref{sec:correlation}  for both datasets. 

\subsubsection{IMD Data}
The pattern correlation and RMSE obtained from the IMD data is shown in Figure 10. It is seen that pattern correlation      worsens from lead day 1 to 2. Further, the pattern correlation shows large variations across the Indian region. The best correlations are noted over the west coast and monsoon trough region, while the lowest values are noted over the northern regions. The 2 lead days' patterns are reasonably correlated over the Western Ghats and monsoon trough region (~0.8 on lead day 1 and ~0.6 for lead day 2).  However, the model fares poorly over the parts of Himalayas regions and Rajasthan. 

Over these regions, the pattern correlations deteriorate quickly after lead day one (Ld1) and reach below 0.4 on lead day 2 (Ld2). One plausible reason behind the poor correlation over these regions might be the sparse density of IMD stations (as mentioned in \cite{pai2014development}). We also computed the 3 days average skill of the models that is average of Ld1,Ld2 and Ld3 forecasts. These plots are shown in the bottom panel of Figure 10.  This 3 days average shows similar skill as in case of Ld1 and Ld2.

\begin{figure}[ht]
	\centering 
	\includegraphics[scale=0.46]{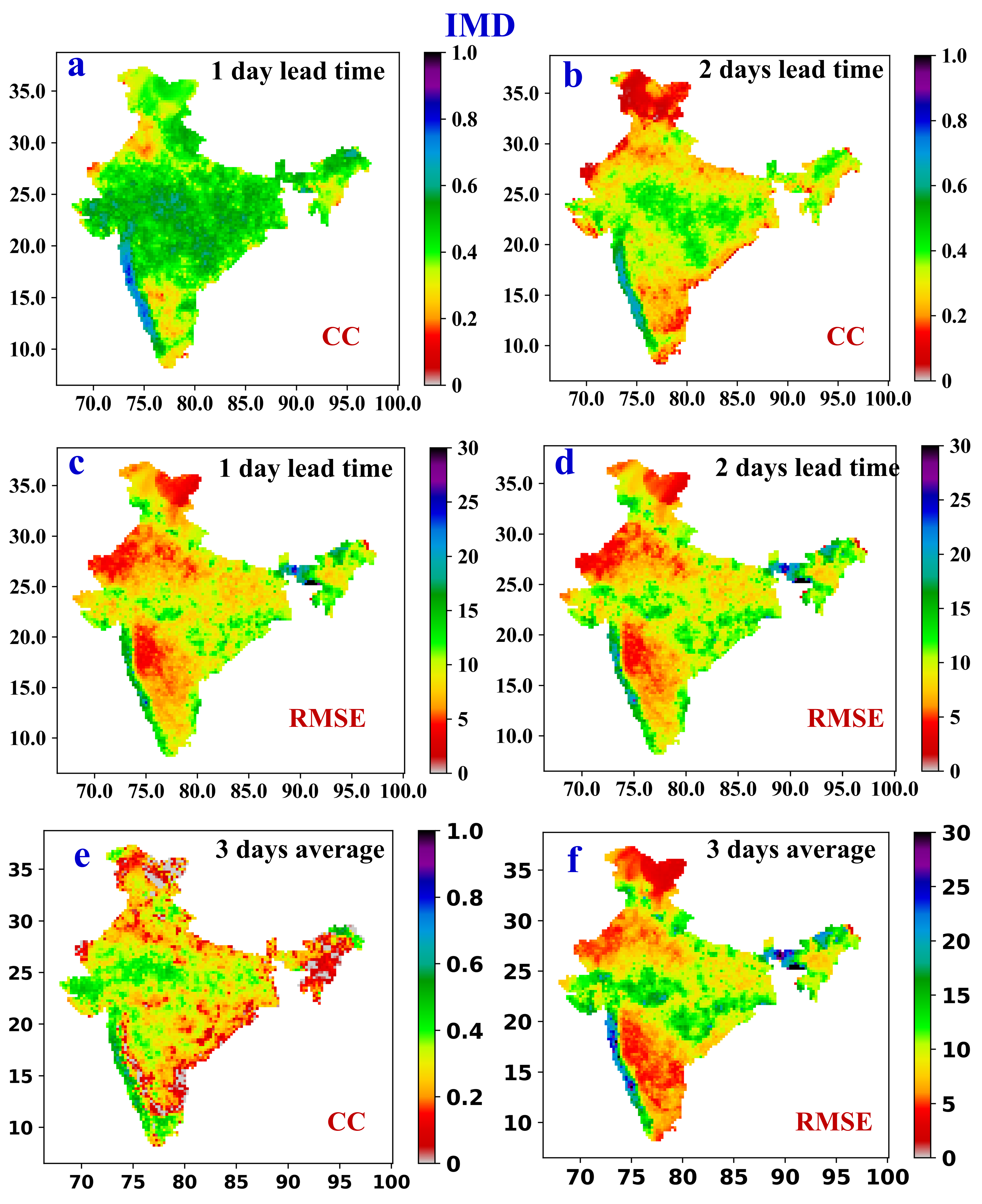}
	\caption{ Correlation (panels a \& b) and RMSE (panels c \& d) of Ld1 and Ld2 for IMD data. }
	\label{fig:cc_rmse_imd}
\end{figure}

Nevertheless, the model reasonably captures the variability in the short term.  The RMSE for the Ld1 and Ld2 are shown in the lower panel of Figure 10. The RMSE is considerably low in most regions except in some parts of the North East area (around the Sikkim region). Relatively higher values of RMSE can also be seen in the Western Ghats area for both Ld1 and Ld2. The model’s performance is comparable to state of art numerical weather prediction models  \cite{rao2019monsoon, rajeevanindia}. 

\subsection{TRMM data}
the same formula given in equation 4.  In this case, with increasing lead time, the pattern correlation decreases significantly as shown in Figure 11 (panels a \& b). The model requires improvements to capture the rainfall for TRMM data better. One possible improvement can be to use multivariable input for training. The RMSE for Ld1 and Ld2 for TRMM data are presented in lower panels (panels c and d) of Figure 11. It is noted that the Western Ghat area has higher CC and lower RMSE. 
\begin{figure}[ht]
	\centering 
	\includegraphics[scale=0.46]{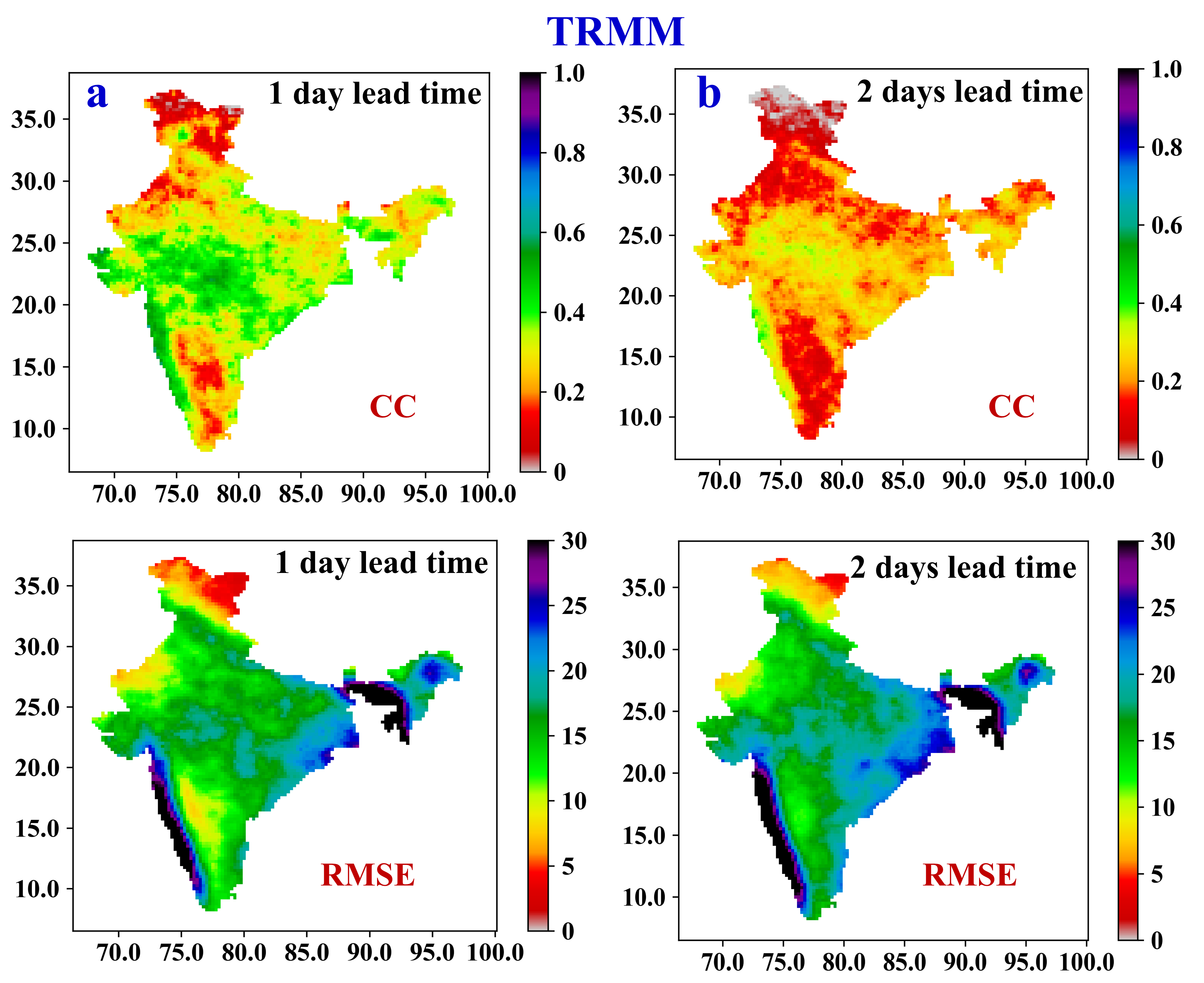}
	\caption{ Correlation (panels a \& b) and RMSE (panels c \& d) of Ld1 and Ld2 for TRMM data. }
	\label{fig:cc_rmse_trmm}
\end{figure}

\subsubsection{Homogeneous regions of IMD data}
The pattern correlations for homogeneous regions show a similar trend as in the entire Indian territory, which means it deteriorates after day 1. Figure \ref{fig:cc_west_central} depicts the pattern correlations for West Central (panel a) and Central North-East (panel b) regions. A better CC was obtained in the Central NE area for the Ld1.For the second day lead time, the CC falls quickly in both areas; however, RMSE in West Central does not make much difference as shown in Figure \ref{fig:cc_west_central}.

\begin{figure}[ht]
	\centering 
	\includegraphics[scale=0.40]{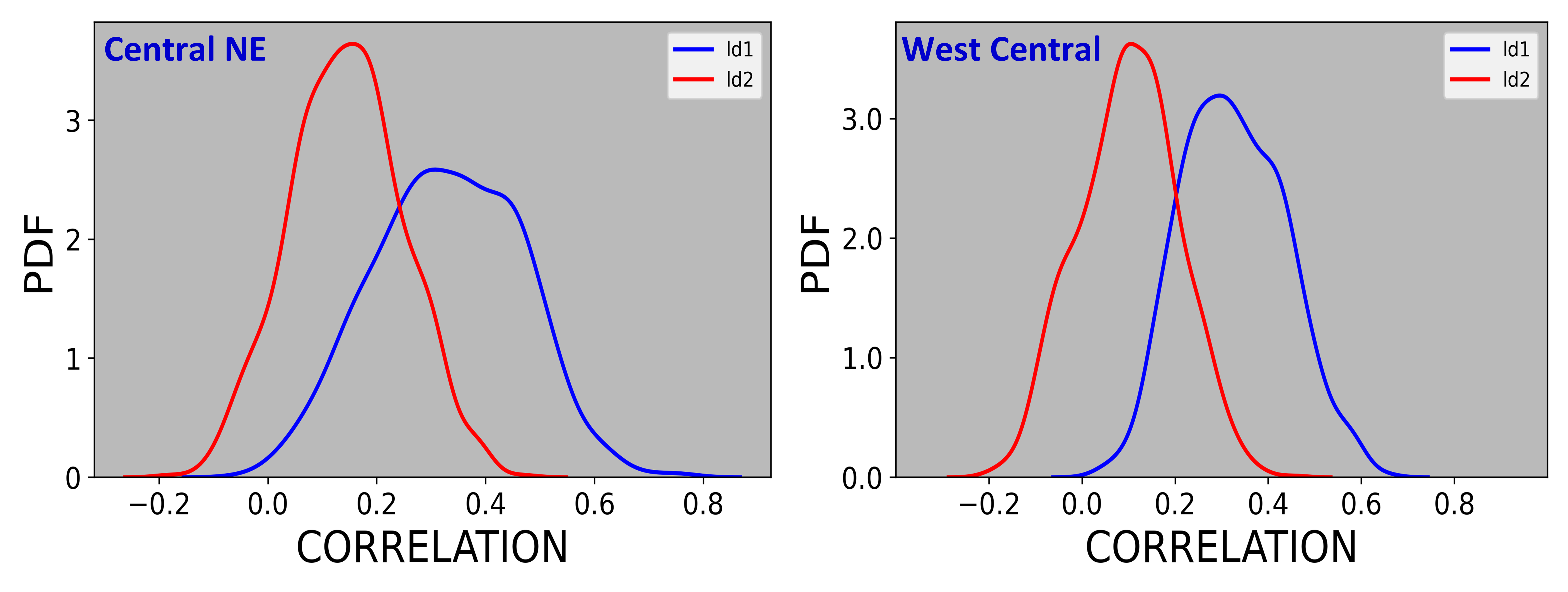}
	\caption{Comparison of the pattern correlation for Central North East (panel (a) ) and West Central ( Panel (b) ) for 2 days  lead time. A comparison for pdfs of observed and predicted rainfall values are shown in panel (c). }
	\label{fig:cc_west_central}
\end{figure}

Moreover, the comparison of pdfs for observed and predicted rainfall for the entire Indian landmass (panel (c)  of figure 12) indicates that both Ld1 and Ld2 forecast underestimate the heavy rainfall events,  a common problem for most of the rainfall forecast models. Further, we calculated the PDF of RMSE for Ld1 and Ld2 of heavy rainfall in these regions. The heavy rainfall days were selected by taking only those days in which atleast 10 percent of the grid points in the homogeneous region had more than 95 percentile rainfall value. A comparison of PDFs is provided in Figure \ref{fig:cc_west_central}. The Ld1 RMSE is found to be less than Ld2 for three homogeneous regions, namely, Central NE, West Central and North East. There was no difference in RMSE between Ld1 and Ld2 forecasts was found for the other three regions.      
\begin{figure*}[ht]
	\centering 
	\includegraphics[scale=0.46]{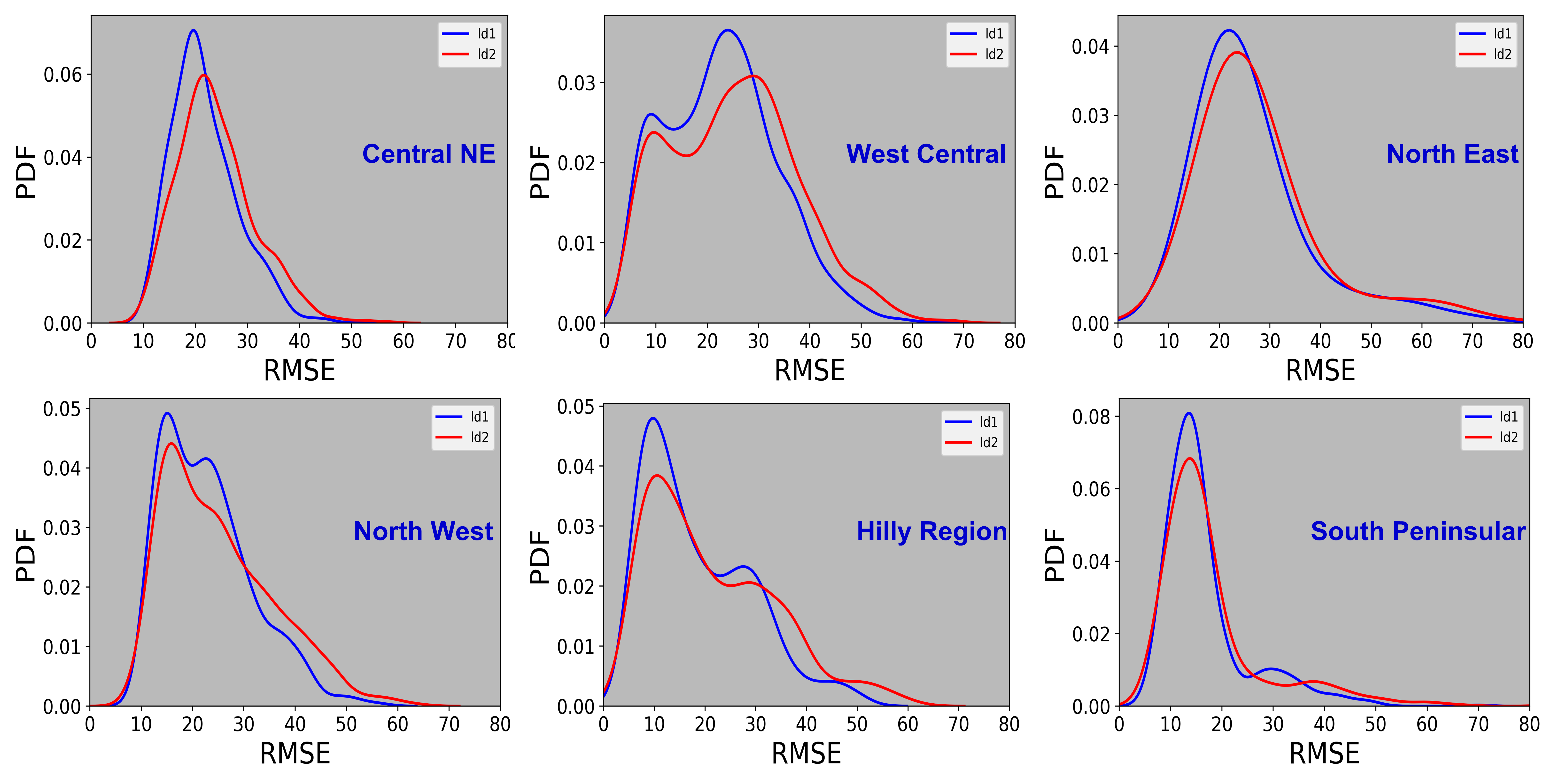}
	\caption{A comparison of PDF of RMSE calculated for lead day 1 (Ld1) and lead day 2 (Ld2). The RMSE for Ld1 is found to be less than Ld2 in the upper panels representing three regions: Central NE, West Central and North East. }
	\label{fig:cc_west_central}
\end{figure*}

\subsection{Calculation of Receiver Operating Characteristics (ROC) curve}
Another skill   metric  calculated for homogeneous regions is receiver operating characteristics (ROC),  defined in section II. A description of the application of the same method is provided in Caren Marzban \cite{marzban2004roc}, highlighting it as a measure of classification performance. 

In our study, we have used a simple skill verification method as well as category (or threshold) based classifier verification. We calculated TPR and FPR (equation 7.1) for rainfall values in all six regions after binning the rainfall in different categories. The categories are determined based on minimum and maximum rainfall values and then slices them in 1mm intervals. Category-wise comparison indicates the skill of different rainfall bins, thus giving an idea on how the skill varies in different rainfall categories. Comparisons of these rates for all regions are provided in Figure 14. The blue dots indicate Ld1 forecast and orange dot represent the Ld2 forecast. The blue curve has larger Area Under the Curve (AUC) values, consistent with the correlation values (i.e. skill of Ld1 greater than the skill of Ld2) for these regions. The North West region does not show much difference in Ld1 and Ld2 skill.

\begin{figure*}[ht]
	\centering 
	\includegraphics[scale=0.46]{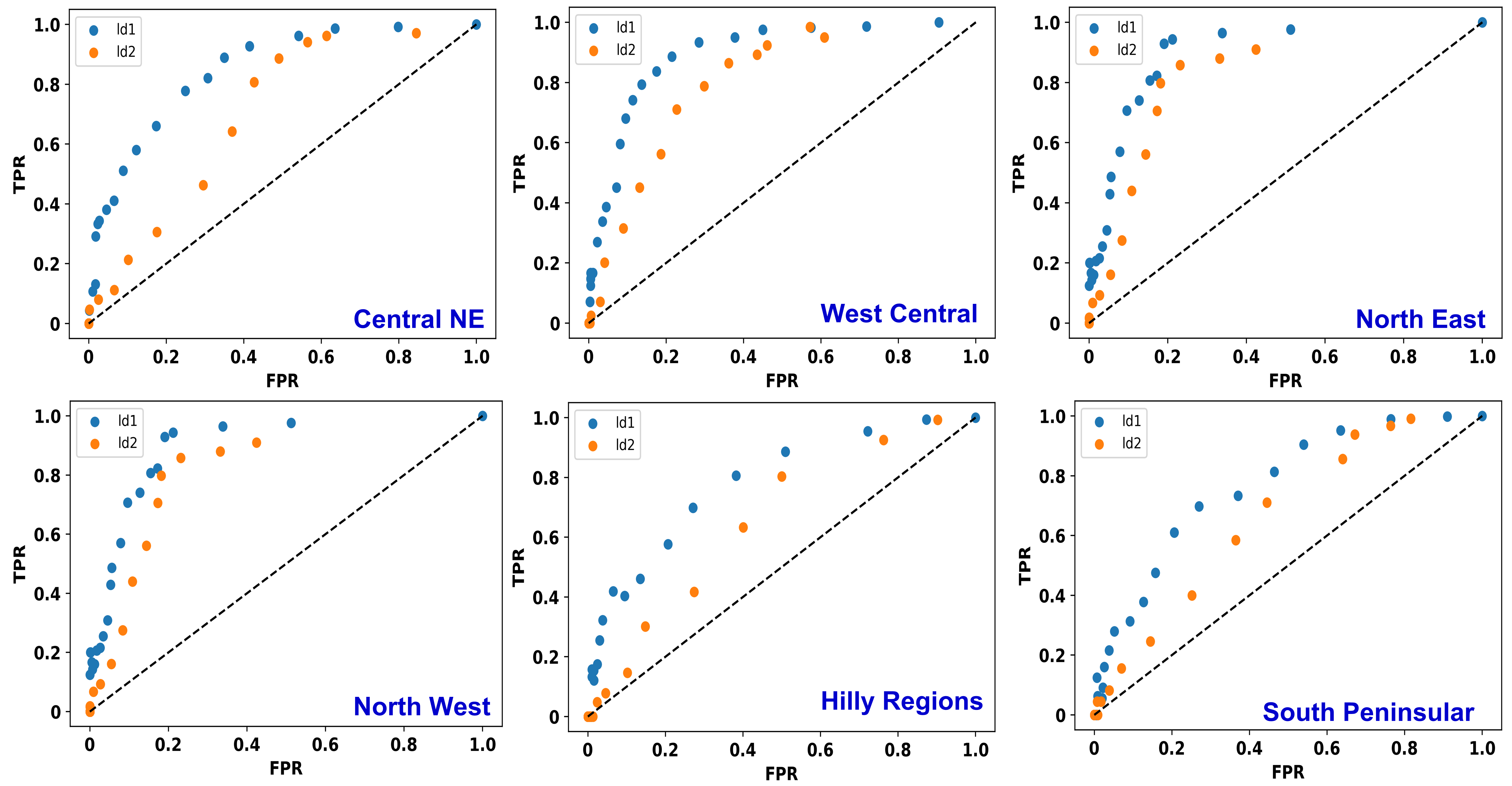}
	\caption{Comparison of ROC skill for different homogeneous regions. The Ld1 TPR is better for three regions: Central NE, West Central and North West. }
	\label{fig:ROC}
\end{figure*}

\section{Conclusion}

This study focused on implementing a deep learning model, namely, ConvLSTM for the short-range forecasting of the ISMR. ConvLSTM based models have been used for short-range forecasts elsewhere with some success. The proposed model is a proof-of-concept which can capture the spatio-temporal structure of the forecast data. It was employed on two different observation datasets, namely, IMD and TRMM datasets.  

The convolution operation is not well-defined in the literature when we do not have data over a certain spatial domain. The IMD data, for example, do not have values over the ocean. Such sharp gradient at land-sea boundaries can be potentially problematic for convolution operation due to the absence of data. We applied an efficient approach and tackled the undefined values (i.e., grids having no data).  For which the data were first transformed from real space to exponential space. The model training was done in exponential space allowing rainfall values to span $(1, \infty)$; replacing the NaN values with ‘0’, as ‘0’ was no more significant value in this space. 

The model-produced forecast shows reliable skill with observations (the ground truth); however, up to 2 days lead time only.  The efficiency quickly goes down after that, as seen in the pattern correlations. A low correlation is seen at the northern and North Western parts along the east coast of India. The forecast is also done separately for homogenous monsoon regions described in the \cite{kothawale2017monthly}. In this case, the area-averaged correlation for 5 years’ time series is found to be reasonably good, and the RMSE for this data is significantly low. However, the pattern correlations again fall quickly after 1 day lead time. The forecast obtained from this deep learning model is comparable with the same is from state of the art dynamical models such as provided in \cite{mukhopadhyay2019performance}. The forecast skill was also analysed using the ROC curve for homogeneous regions. The ROC analysis was found to be consistent with correlation. We note that the present model reasonably captures the widespread precipitation but still have issues with localized events which might be related to the fact that large scale organized systems have more lifetime and spatial scale which can be captured based on the single variable model attempted here. While the localized extremes are often of short duration and do not have enough memory with them to be taken for the next day when dealing with daily data. Therefore it is still a challenge even for state-of-the-art NWP models to predict such events.

\section{DISCUSSION AND FUTURE WORK}

This work is a  demonstration of deep machine learning-based algorithms for weather forecasting using only a single variable, which is probably a reason for the steep fall in the efficiency of forecasts after 2 days. Thus, the model, in the present form has limitations. We could not compare this with other models due to limited availability of short-range monsoon forecast models based on 2D convolution and RNN based techniques. However, it is noted that the two day lead predictions of this model compare reasonably against the global forecast system (GFS) T574L64 ($\approx$5 km), adopted from National Centers for Environmental Prediction (NCEP), and tested by the IMD during the 2010s \cite{durai2014prediction}. The study reported that areas of negative mean errors spread over most parts of the country from the lead day-2 onwards. With the adoption of higher resolution and improved GFS T1534 ( $ \approx$ 12.5 km), the efficiency of short range operational forecasts have increased \cite{mukhopadhyay2019performance}. 

It has been reported that GFS T1534 has much improved skill in moderate (15.6 - 64.5 mm day - 1) rainfall categories while there is underestimation for the heavy to very heavy (64.5 - 204.05 mm day - 1) rainfall. Also the extremely heavy rainfall categories are only better on the shorter lead times. This ensemble based state-of-the-art forecasting is efficient but resource intensive       and has issues as discussed. We state that our model is proof of concept model to utilize ConvLSTM based architecture for forecasting of ISMR. We acknowledge that there could be other models also which could be successful in predicting rainfall. We are also experimenting to improve the model.  The main purpose of the current study is to introduce a model which can be used to forecast monsoon rainfall on short scale.

Multivariate learning is essential to capture the low-frequency variability of rainfall as low-frequency sub-seasonal waves are convectively coupled waves with moisture, the surface low-pressure, and wind. Hence, the potential variables which can be used in further studies are sea level pressure, air temperature, and humidity. The technique can be improved by using more layers in the training and the tuning of hyper parameters.

The custom loss function used for TRMM data can also be experimented on the IMD dataset to improve the data training.  Other architectures that can be tried out is the Gated Recurrent Unit (GRU) \cite{shi2017deep}, a newer generation of RNN and is more straightforward than LSTM. Another modification that can be done to handle datasets like IMD with NAN values is to generalize the convolution operator to act on irregular shapes \cite{vialatte2016generalizing, pasdeloup2017convolutional}.  This model has potential to be utilized in   short-range forecasting of monsoon precipitation, fire prediction and heat/cold wave forecasting. One can develop a multi-model ensembles using different architectures.

%\appendices
%\section{Proof of the First Zonklar Equation}
%Appendix one text goes here.
%
%% you can choose not to have a title for an appendix
%% if you want by leaving the argument blank
%\section{}
%Appendix two text goes here.

% use section* for acknowledgment
\section*{Acknowledgment}

The IITM Pune is funded by the Ministry of Earth Science (MoES), the Government of India.  This work was done using HPC facilities Pratyush (http://pratyush.tropmet.res.in/) provided by MoES at IITM Pune. 

% Can use something like this to put references on a page
% by themselves when using endfloat and the captionsoff option.
\ifCLASSOPTIONcaptionsoff
  \newpage
\fi

% references section

% can use a bibliography generated by BibTeX as a .bbl file
% BibTeX documentation can be easily obtained at:
% http://mirror.ctan.org/biblio/bibtex/contrib/doc/
% The IEEEtran BibTeX style support page is at:
% http://www.michaelshell.org/tex/ieeetran/bibtex/
\bibliographystyle{IEEEtran}
% argument is your BibTeX string definitions and bibliography database(s)
\bibliography{IEEEabrv,convlstm_bibtex}
\end{document}